\documentclass[twocolumn,showpacs,preprintnumbers,amssymb,prd]{revtex4}

\usepackage{graphicx}
\usepackage{dcolumn}
\usepackage{bm}

\begin{document}
\input{epsf}

\title{Addendum to ``Coherent radio pulses from {\small GEANT}
generated electromagnetic showers in ice''}

\author{Soebur Razzaque,$^1$ Surujhdeo Seunarine,$^2$ Scott
W. Chambers,$^3$ David Z. Besson,$^3$ Douglas W. McKay,$^3$ John
P. Ralston,$^3$ and David Seckel$^4$}

\affiliation{$^1$Department of Astronomy \& Astrophysics, Penn State
University, University Park, Pennsylvania 16802}

\affiliation{$^2$Department of Physics and Astronomy, University of Canterbury,
Private Bag 4800, Christchurch, New Zealand}

\affiliation{$^3$Department of Physics \& Astronomy, University of
Kansas, Lawrence, Kansas 66045}

\affiliation{$^4$Bartol Research Institute, University of Delaware,
Newark, Delaware 19716}

\begin{abstract}
We reevaluate our published calculations of electromagnetic showers
generated by {\small GEANT 3.21} and the radio frequency pulses they
produce in ice.  We are prompted by a recent report showing that
{\small GEANT 3.21}-modeled showers are sensitive to internal settings
in the electron tracking subroutine.  We report the shower and
pulse characteristics obtained with different settings of {\small
GEANT 3.21} and with {\small GEANT 4}. The default setting of electron
tracking in {\small GEANT 3.21} we used in previous work speeds up the
shower simulation at the cost of information near the end of the
tracks.  We find that settings tracking $e^-$ and $e^+$ to lower
energy yield a more accurate calculation, a more intense shower, and
proportionately stronger radio pulses at low frequencies.  At high
frequencies the relation between shower tracking algorithm and pulse
spectrum is more complex.  We obtain radial distributions of shower
particles and phase distributions of pulses from 100 GeV showers that
are consistent with our published results.
\end{abstract}

\date{\today}
\maketitle

In a recent paper \cite{soeb} we reported the results of an extensive
study of the radio frequency pulse emitted by an electron-induced
electromagnetic shower.  Application to radio signal detection of
ultrahigh energy electron neutrino interaction in ice provided our
motivation \cite{radhep}.  We used the standard particle physics
detector-simulation package {\small GEANT 3.21} (G3 afterwards)
\cite{geantman} with default settings to generate the showers.  Our
total track-lengths of electrons and number of electrons at shower
maximum were 68\% of those in the paper by Zas, Halzen and Stanev
(ZHS) \cite{zhs}.  Our pulse height at the Cherenkov angle at 1 GHz
was 72\% of the ZHS result.  Our study of sensitivity of shower
development to reasonable changes in input cross sections showed no
effects large enough to explain the difference between G3 and ZHS
shower profiles and track lengths.

The source of the bulk of the discrepancy was recently presented by
Alvarez-M\~uniz, Marqu\'ez, V\'azquez and Zas \cite{jaime}. They
noticed that the default setting of a parameter called ``{\small
ABAN}'' in the electron tracking subroutine of G3, the setting used in
our studies, gives shorter total, projected and weighted track lengths
and lower particle yield at shower maximum than {\small GEANT 4} (G4
afterwards) or G3 with a choice of tracking rules different from the
default.  Their results for track-length and shower profile from the
latter simulations are  90\% of those of the original ZHS work
for these quantities and in closer agreement than G3 showers with the
default setting \cite{soeb, jaime}.

Track lengths are one indicator of the strength of electric field
pulses to be expected from the shower.  However the electric field at
a given angle and frequency depends on a combination of many factors.
We make a comparison among the shower {\it and} electric field pulse
results from G3 (with several different combinations of electron
tracking rules) and G4.  We report results for shower profiles, total
track lengths, energy loss ($dE/dx$) and radial distribution of the
number of electrons minus the number of positrons around the shower
maximum.  We tabulate the G3, G4 and ZHS track-lengths and yields at
shower maximum for ease of comparison of shower results.  Regarding
the electric field in the Fraunhoffer regime, we report radio pulse
strength at the Cherenkov angle and the phase behavior of the radio
pulse both on and off the Cherenkov peak.  We find that the particle
yield and charge imbalance are larger with improved tracking.  At low
frequencies the pulses emitted at the Cherenkov angle are
correspondingly stronger as well.

{\it Electron tracking in GEANT:} The subroutine responsible for
tracking electrons in G3 is called {\small GTELEC}. In its default
setting, {\small ABAN=1}, if the electron range is too short to reach
the boundary of a detector volume, and the distance to its next
bremsstrahlung emission is greater than its range, the electron's
energy is deposited in the volume and it is stopped.  This is done
regardless of the value of the electron kinetic energy threshold
parameter {\small CUTELE} set by a user in the input card file. This
was presumably made the default to increase the calculation Al speed of
high energy particle physics detector simulations in non-sensitive
volumes, where detailed tracking was not needed \cite{aban}. With the
default setting, however, track lengths are reduced significantly,
with a consequent reduction in the calculated radio Cherenkov
emission. This is what we observed in our previous study \cite{soeb},
where we compared results from G3 with default setting to the
results of \cite{zhs}.

In practice, it seems for applications such as ours the preferred
\cite{michel1} setting is {\small ABAN=0}, which tracks all electrons
down to the threshold set by the user. The setting {\small ABAN=2}
allows the user to track electrons precisely (similar to {\small
ABAN=0} setting) in sensitive volumes or not so precisely in the
non-sensitive volumes \cite{isvol}.  ``Sensitive''
vs. ``non-sensitive'' settings are distinguished by the value of a
step parameter called ``{\small STMIN}'' in the tracking medium
parameter list, where ``sensitive'' produces many more steps per track
than ``non-sensitive''.  G4 has range cuts that correspond to energy
thresholds for the production of particles.  A particle, once
produced, is tracked down to rest regardless of its production
threshold.

{\it Results:} We expand on the study reported in Ref. \cite{jaime},
where showers produced with G3 {\small ABAN=1}, G3 {\small ABAN=2}
with ``sensitive'' volume, G4 and ZHS are compared.  We run G3 for
four different tracking choices, namely {\small ABAN=0}, {\small
ABAN=1}, {\small ABAN=2} sensitive volume (S) and {\small ABAN=2}
non-sensitive volume (NS) settings, to create showers in ice. We run
G4 for comparison as well.  We then repeat our study \cite{soeb} of
the key features of the showers and the consequent key features of the
radio pulses: pulse angular distributions, phase distributions and
frequency spectra.

{\it 1) Shower Characteristics:} First we investigate the important
issue of total track lengths in Table~\ref{tab:tracklengths}.  We note
that the track lengths for {\small ABAN=2} (NS) setting are similar to
the default {\small ABAN=1} setting; the {\small ABAN=2} (S) setting
used in Ref. \cite{jaime} gives results similar to the preferred
{\small ABAN=0} setting and both are similar to the results from G4.
The agreement between track lengths produced by these latter three
GEANT versions and those produced by the ZHS code improves as one
compares total, total projected $e+p$ and total projected $e-p$ track
lengths.  Comparing the projected $e+p$ and $e-p$ track lengths, we
see that the loss of track length for {\small ABAN=1} is more
significant for electrons than for positrons.  We expect this to be
the case, since a large part of the electron excess comes from low
energy electrons swept into the shower by Compton scattering and
$\delta$-ray production.
\begin{table*}
\caption{\label{tab:tracklengths} Track length and particle yield results 
from an average of 100 GeV electron induced shower using different Monte 
Carlo shower codes.  The error bars 
correspond to error in the mean, $s/\surd{N}$, where $s$ is the standard
deviation reported within parentheses and $N$ is the number of showers 
used in each case. }
\begin{ruledtabular}
\begin{tabular}{lc|ccc|ccc}
Monte Carlo & No. of & \multicolumn{3}{c|}{Total track lengths} &
\multicolumn{3}{c}{Particle yield at shower max} \\ 
shower code & showers & Absolute & Projected & Projected 
& Total & Excess & Excess \\
&  & ($e+p$) [m] & ($e+p$) [m] & ($e-p$) [m] 
& ($e+p$) & ($e-p$) & fraction \\
\hline
G3 ({\small ABAN=0})\footnote{Preferred setting (new).} & 20 
& 542.74 $\pm$ 0.08 (0.36) & 455.3 $\pm$ 0.2 (0.9) & 125.0 $\pm$ 2.0 (9.0) 
& 148 $\pm$ 5 (22) & 42 $\pm$ 3 (13) & 28\% \\
G3 ({\small ABAN=1})\footnote{Default setting (old).} & 20 
& 389.51 $\pm$ 0.48 (2.16) & 365.7 $\pm$ 0.5 (2.2) & 76.3 $\pm$ 1.5 (6.7) 
& 111 $\pm$ 7 (31) & 20 $\pm$ 2 (9) &  18\% \\
G3 ({\small ABAN=2}, NS) & 20 
& 389.40 $\pm$ 0.71 (3.18) & 367.0 $\pm$ 0.7 (3.1) & 78.4 $\pm$ 2.8 (12.5)
& 113 $\pm$ 10 (45) & 25 $\pm$ 5 (22) & 22\% \\
G3 ({\small ABAN=2}, S)\footnote{Setting used in Ref. \cite{jaime} 
(similar to {\small ABAN=0}).} & 20 
& 561.05 $\pm$ 0.09 (0.40) & 459.8 $\pm$ 0.3 (1.3) & 125.8 $\pm$ 1.9 (8.5)
& 148 $\pm$ 8 (36) & 41 $\pm$ 3 (13) &  28\% \\
G4 & 100 
& 572.58 $\pm$ 0.04 (0.40) & 466.3 $\pm$ 0.2 (1.6) & 135.0 $\pm$ 0.8 (7.9) 
& 153 $\pm$ 3 (27) & 45 $\pm$ 1 (13) &  30\% \\
ZHS	   
& 20 & 642.17 $\pm$ 0.06 (0.25) & 516.6 $\pm$ 0.2 (0.9) 
& 135.2 $\pm$ 1.5 (6.8) & 164 $\pm$ 6 (28) & 44 $\pm$ 2 (11) & 27\% 
\end{tabular}
\end{ruledtabular}
\end{table*}

Despite the improved agreement on total track lengths, there are still
significant differences between the G3 {\small ABAN=0}, G3 {\small
ABAN=2}(S), and G4 simulations.  For each code individually, however,
the variance in total track length $36-40$~cm is reasonable. For a
given shower energy $E_{0}$, the total track length $L_{tot}$ is
determined primarily by the rate of ionization energy losses, where
the energy transfer is less than a low threshold $E_t \sim
100$~keV. Then $L_{tot} \approx E_0/(dE/dx)_{ion}$, as expected under
the conditions that a) soft ionization losses are essentially
independent of particle energy and b) hard interactions do not
``remove" energy from the shower. Hard $\delta$-rays create their own
track length and hard $\gamma$'s return their energy to the shower in
pair production and Compton scattering processes. Some energy is lost
from the shower as soft particles fall below $E_t$.  The fluctuations
in track length are thus determined by the fluctuations in the number
of soft particles multiplied by $E_t/(dE/dx)_{ion}$.  Estimating the
number of soft particles by $N_t = E_0/E_t$, the fluctuations in total
track length are of order $\delta L_{tot} \sim E_t/(dE/dx)_{ion}
\sqrt{N_t} =\sqrt{E_o E_t}/(dE/dx)_{ion} = 50$~cm, for $E_0=100$~GeV,
$E_t=100$~keV, and $(dE/dx)_{ion} = 2$~MeV/cm. Although this is a
rough estimate, the observed values of $\sigma$ for $L_{tot}$ are
consistent with expectations for the newer simulations.

This estimate leads to several observations. a) The deviations in
$L_{tot}$ between G3 {\small ABAN=0}, G3 {\small ABAN=2}(S), and G4
are small but significant; this is under study. b) Due to the
default decision in the tracking code to abandon some particles, the
older G3 simulations have an effective value of $E_t\sim 5$~MeV, which
implies $\delta L \sim 3.5$~m, generally consistent with the Table
entries. c) For $e-p$ track length, the fluctuations are dominated by
the competition between pair production and Compton scattering, for
which one may estimate $E_t \sim 20$~MeV and $\delta L \sim 7$~m.

Particle yields at the maxima of longitudinal shower profiles are also
listed in Table~\ref{tab:tracklengths}.  The results from the ZHS
code, G3 with {\small ABAN=0}, {\small ABAN=2} with sensitive volume,
and G4 all agree reasonably well.  Agreement is especially good for
the excess charge, where results agree within errors.  Again, the low
yields from G3 {\small ABAN=1} and {\small ABAN=2} (NS) are consistent
with the interpretation that lost track length is caused by a
deficiency of low energy electrons being scattered into the shower.

Next we consider how individual particle tracks contribute to the
radio pulse.  For low frequencies it can be shown that the
contributions are in phase and the pulse is proportional to the
weighted, projected track length, labeled (e-p) in Table 1. As the
frequency increases, this simple coherence breaks down as described by
two parameters, which we called \cite{soeb} the Cherenkov Phase (CP)
and the Translational Phase (TP).  The CP is related to the phase
coherence within a single track segment, whereas the TP describes the
relative phase among track segments due to their different positions
within the shower, relative to the observer.

In general, we find that the different simulations produce phase
distributions similar to those shown in Fig. 17 of Ref.~\cite{soeb}.
Specifically, the radial charge distribution at shower maximum, shown
to peak at $\rho \simeq 0.5$~cm in Fig.~22 of Ref.~\cite{soeb}, is
reproduced apart from an overall increase in normalization
\cite{buniy}.  Moreover, the radial charge distribution early in the
shower evolution has a core dominated by high energy particles, and
the normalization of this core does not depend on the details of the
electron tracking at low energy.

\begin{figure}
\centerline{\epsfxsize=3.4in 
\epsfbox{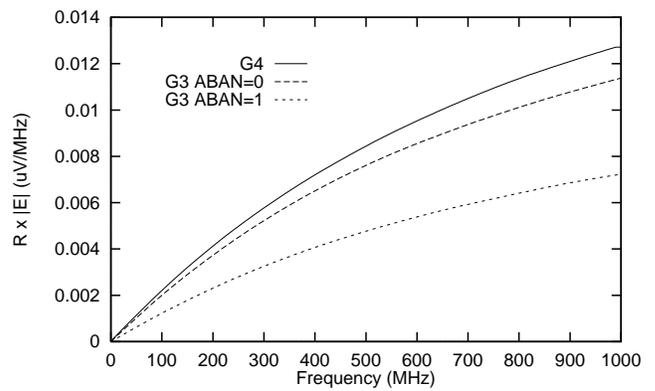}}
\caption{Direct Monte Carlo calculation of the low frequency end of
the spectrum of the average electric field amplitude for twenty 100
GeV showers.  The showers are run using G4 and G3 with both the
preferred {\small ABAN=0} and the default {\small ABAN=1} settings.
The same code was used to calculate electric fields from all the
{\small GEANT} showers.}
\label{fig:lofrqc}
\end{figure}

{\it 2) Radio frequency pulse:} Next we examine how the changes in
shower simulation affect the modeled radio Cherenkov pulses.  At low
frequencies, all particles contribute in phase, so the increase in the
excess track length found in G3 {\small ABAN=0} and G4 should be
reflected by an increase in the amplitudes of the low frequency part
of the spectrum as compared to showers produced with G3 {\small
ABAN=1}.  In Fig.  \ref{fig:lofrqc} we see that this is indeed the
case, where the spectrum $R\times |\vec{E}|$ is shown for an observer
on the Cherenkov cone.

\begin{figure}
\centerline{\epsfxsize=3.4in 
\epsfbox{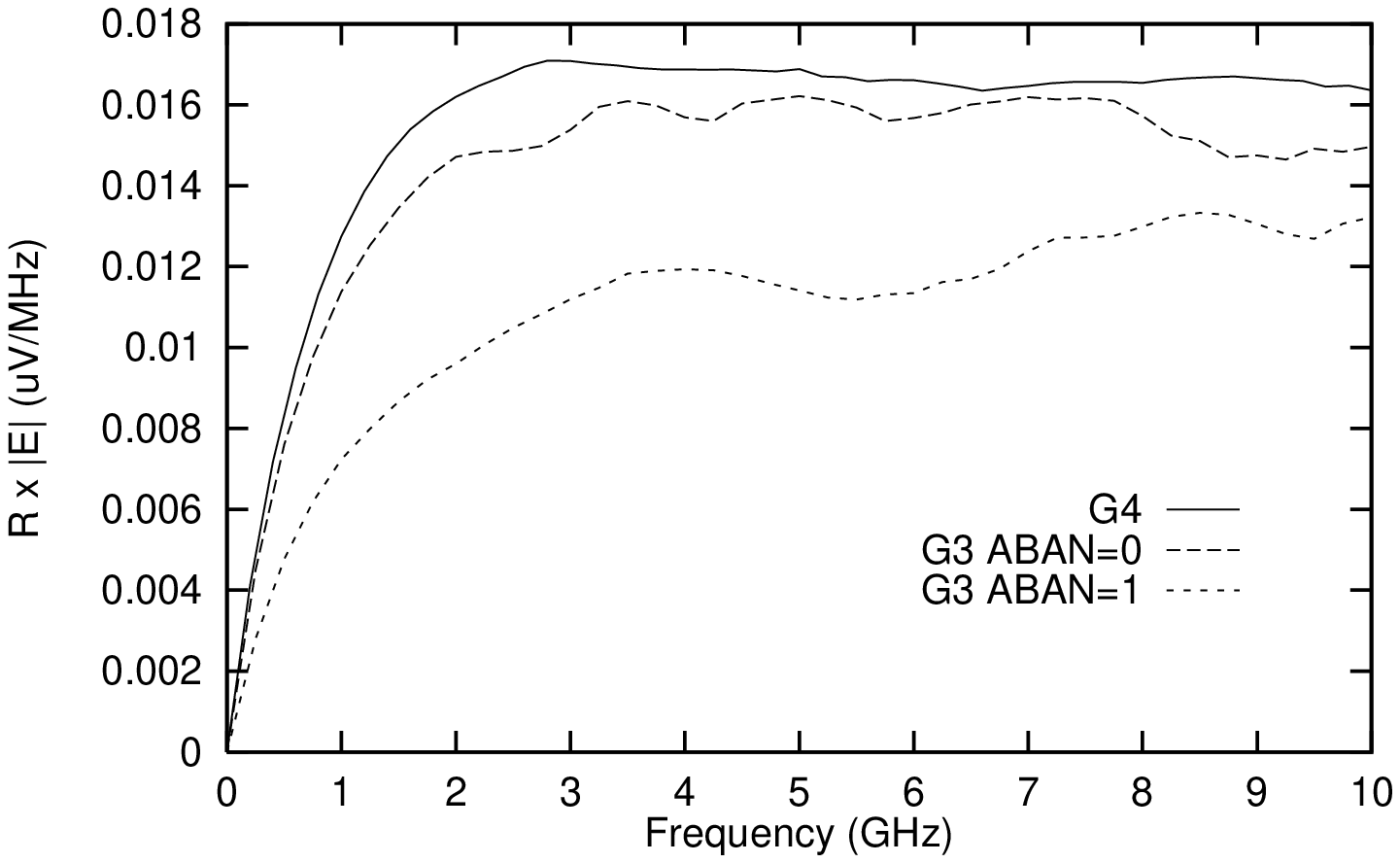}}
\caption{Direct Monte Carlo calculation of the frequency spectrum of
the average electric field amplitude for twenty 100 GeV showers.  The
showers are run by G4 and G3 with both the preferred {\small ABAN=0} and the
default {\small ABAN=1} settings.  The same code was used to calculate
electric fields for all the {\small GEANT} showers.}
\label{fig:frqc}
\end{figure}

In Fig.\ref{fig:frqc} we extend the frequency scale out to 10 GHz,
which suffices to show clearly a transition from low to high frequency
for the same sets of simulations.  In this range, the spectra for the
G3 {\small ABAN=0} and G4 showers flatten out, but the G3 {\small
ABAN=1} spectrum grows slowly, so that the difference in their
normalizations, relative to G4, is less than 20\% for $f = 10$~GHz.
We infer that in this region the spectra are not simply described by
the phase coherent sum of amplitudes from a large number of low energy
electrons.  We note two effects that could be important in this
region.  First, a relatively small number of high energy particles
near the shower axis could contribute strongly to the spectrum in this
region.  Since this part of the shower evolution is not sensitive to
low energy electron tracking, we may see the G3 and G4 results
converging.  This trend continues at higher frequency.  Second, a
large fraction of the track-segment amplitudes are no longer coherent.
For that fraction, the magnitude of $\vec{E}$ increases as a random
walk as the track length increases.  In this case, one would expect
the amplitude to increase as the square root of the number of track
segments, rather than linearly, which reduces the ratio of the pulse
from the G4 generated shower to that from {\small ABAN=1} as the
frequency increases.  This is seen clearly in Figs. 1 and 2.  Given
that total and projected track lengths are proportional to energy
\cite{soeb,zhs}, the effects just outlined should be relevant when
comparing pulses from showers of different energies using the same
simulation.

\begin{figure}
\centerline{\epsfxsize=3.4in 
\epsfbox{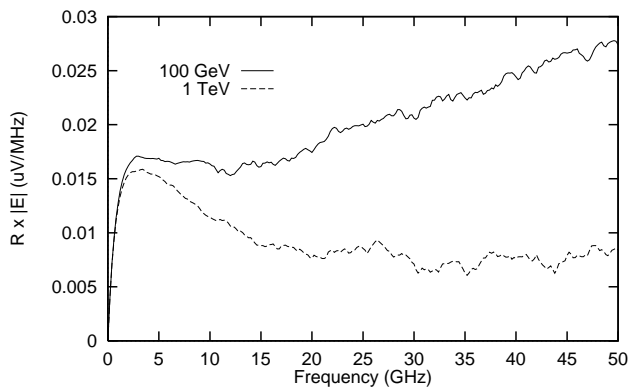}}
\caption{Direct Monte Carlo calculation of the frequency spectrum of
the average electric field magnitude for one hundred 100 GeV showers 
and ten 1 TeV showers run by G4. The 1 TeV spectrum is divided by a
scaling factor of 10.}
\label{fig:G450}
\end{figure}

Figure \ref{fig:G450} shows a comparison between pulse spectra from 1
TeV showers and 100 GeV showers, both generated by G4.  The spectra
are extended up to 50 GHz.  For frequencies above 15 GHz, the results
are fully consistent with Fig. 25 of Ref.~\cite{soeb}.  At low
frequencies, below a few GHz, the results are consistent as well,
after accounting for the change in normalization between G4 and G3
{\small ABAN=1}.  At 50 GHz, the TeV average, scaled down by a factor
of 10, is smaller by roughly a factor $3 \sim \sqrt{10}$ than the 100
GeV average, possibly hinting that the ``random walk'' increase of
$\vec{E}$ with energy, and therefore total track length, is setting
in.  At the same time, it seems clear that a combination of effects
must be in play for intermediate frequencies.

Since the flat spectra observed above several GHz contrast with the
sharp falloff expected above the 1-2 GHz scale \cite{zhs, avz00}, we
referred to this phenomenon as an ``extended coherence regime'' in
[1].  This is different from the use of coherence to refer to the
rapid rise of the electric field strength as frequency increases in
the region where frequency is low compared to the scale set by the
peak at 0.5 cm in the shower radial distribution.

Summarizing, we see that the improved electron tracking performed by
G4 or G3 {\small ABAN=0} results in proportionately increased radio
Cherenkov spectra at frequencies below a few GHz.  At high frequencies
the situation is more complicated.  We are conducting further studies
at a variety of energies and including showers initiated by
$\gamma$-rays to identify the determining factors.


{\it Summary and further discussion of results:}

Motivated by the observation in \cite{jaime} that the default setting
for electron tracking in G3 stops many electrons prematurely, we have
reviewed the calculations reported in\cite{soeb}. In \cite{soeb} we
used G3 with default setting and found a persistently smaller shower
and a weaker electromagnetic pulse at low frequencies than those
reported in \cite{zhs}.  We find that this discrepancy largely
disappears when electrons are tracked more completely in G3 or when
using G4 generated showers, similar to results reported in
\cite{jaime}.  Figure 1 shows that the strength of the pulses at low
frequency clearly reflect the track lengths and particle yield at
shower maximum.  The high frequency behavior is more complex and is
under further study.

The particles in the electromagnetic showers in a uniform medium, ice
in the case of interest here, should be tracked as long as they still
contribute to the Cherenkov signal.  This is the case, for the ZHS, G4
and G3 simulations with the tracking option {\small ABAN=0} or {\small
ABAN=2} with sensitive volume; the results from all these simulations
agree at the 10\% level.  There is physics, namely significant
emission, that is lost if the tracks are stopped prematurely and their
energy dumped into the medium.  Though choosing the default ({\small
ABAN=1}) setting affects only the low energy tail of the shower, below
the few MeV range, the population of particles (predominately
electrons) is so high at the low energy end that the field
contribution cannot be ignored.

For applications of the shower and pulse codes to neutrino detection
experiments, the difference is obviously important.  For example,
below 1 GHz where currently deployed antennas are sensitive, the
expected radio emission pulse is 1.3 to 1.4 times higher using the
G3 or G4 code with particles tracked to threshold compared to the
truncated showers produced by G3 with the {\small ABAN=1} setting.
The more correct treatments that follow the particles to lower energy
produce significantly larger fields on the Cherenkov cone.  This
increases the effective volume estimates and leads to higher sensitivity
estimates for cosmic neutrino detection systems like the Radio Ice
Cherenkov Experiment (RICE) \cite{ilya} and the recently approved
balloon-borne experiment ANITA \cite{anita}, both using the Antarctic
polar ice cap as a neutrino target.

{\it Acknowledgements:} We thank J. Le Vaillant and J. Adams for
help running {\small GEANT} simulations. We acknowledge helpful
communications with M. Maire, J. Alvarez-Mu\~niz, E. Marqu\'es and
E. Zas. Work was supported in part by the NSF, the DOE and the
University of Kansas General Research Fund. S. Seunarine was supported
by a Marsden grant. S. Razzaque was partially supported by NSF
AST0098416 grant.


\begin{thebibliography}{srt}

\bibitem{soeb} S. Razzaque et al., Phys. Rev. D {\bf65}, 103002
(2002).

\bibitem{radhep} D. Besson in {\it RADHEP 2000}, Los Angeles,
D. Salzberg and P. Gorham Eds., AIP Conference Proceedings, Vol. 579,
pg. 157 (2001). 

\bibitem{geantman} R. Brun et al., {\small GEANT 3} Manual, CERN
Program Library Long Writeup W5013, 1994.

\bibitem{zhs} E. Zas, F. Halzen and T. Stanev, Phys. Rev. D {\bf45},
362 (1992).

\bibitem{jaime} J. Alvarez-Mu\~niz, E. Marqu\'{e}s ,R.  V\'{a}zquez
and E. Zas, astro-ph/0206043.

\bibitem{aban} For historical reasons, the {\small ABAN} choices and
their application are not described in the user's manual \cite{geantman}.

\bibitem{michel1} We thank Michel Maire for communication about the
electron tracking treatments in {\small GEANT 3.21}.

\bibitem{isvol} See parameters {\small ISVOL, DEEMAX} and {\small
STMIN} under {\small CONS200} section of the {\small GEANT} manual
\cite{geantman}.

\bibitem{buniy} R. Buniy and J. Ralston, Phys. Rev. D {\bf65}, 016003
(2002).

\bibitem{avz00} J. Alvarez-Mu\~niz, R.  V\'azquez and E. Zas, Phys. Rev. D
{\bf 62}, 063001 (2000).

\bibitem{ilya} I. Kravchenko et al., Astroparticle Physics {\bf19}, 
15 (2003); I. Kravchenko et al., astro-ph/ 0206371 (Astroparticle
Physics, in press).

\bibitem{anita} P. Gorham, in {\it Ultra High Energy Particles from
Space}, Aspen (2002). 


\end{thebibliography}
\end{document}